\documentclass[a4paper,aps,twocolumn,nofootinbib]{revtex4}
\RequirePackage[colorlinks,hyperindex]{hyperref}
\RequirePackage[english]{babel}
\RequirePackage[latin1]{inputenc}
\RequirePackage[T1]{fontenc}
\RequirePackage{mathrsfs}
\RequirePackage{amsmath}
\RequirePackage{amssymb}
\RequirePackage{amsbsy}
\RequirePackage{color}
\RequirePackage{bm}
\hypersetup{colorlinks=true,breaklinks=true,urlcolor=blue,linkcolor=red}
\begin{document}
\title{\bf{Weyl and Majorana Spinors as Pure Goldstone Bosons}}
\author{Luca Fabbri\footnote{fabbri@dime.unige.it}}
\affiliation{DIME, Sez. Metodi e Modelli Matematici, Universit\`{a} di Genova,
Via all'Opera Pia 15, 16145 Genova, ITALY}
\date{\today}
\begin{abstract}
\textbf{Abstract.} General spinors in polar form display a structure that is made up by real scalar degrees of freedom plus six components which can be recognized as Goldstone bosons: in the present paper we show that of all singular spinors, Weyl and Majorana spinors have no real degree of freedom and so that they can be interpreted as pure Goldstone states.
\end{abstract}
\maketitle
\section{Introduction}
In contemporary mathematical physics, the role of the spinorial fields is essential. From a general perspective, a spinor field is classified by means of bi-linear quantities: the scalar and pseudo-scalar, the vector and axial-vector, and the tensor \cite{L,Cavalcanti:2014wia,HoffdaSilva:2017waf,daSilva:2012wp,Ablamowicz:2014rpa,Rodrigues:2005yz}. According to this classification, the spinors are split in six classes: if scalar and pseudo-scalar are not both identically vanishing they are \emph{regular} (such a class splits into further three sub-classes, but we will not need such a deep categorization in the following) while if both scalar and pseudo-scalar identically vanish they are \emph{singular} and they split into further three sub-classes: for the case where no other bi-linear is identically zero they are called \emph{flag-dipoles}, when the axial-vector is equal to zero they are called \emph{flagpoles} and when the tensor is equal to zero they are called \emph{dipoles}. Regular spinors are Dirac spinors while singular spinors are less known: flag-dipoles have yet to receive a thorough interpretation \cite{Vignolo:2011qt,daRocha:2013qhu}, and although flagpoles are the Majorana spinors and dipoles are the Weyl spinors neither finds place in the description of nature yet. A special case of Majorana spinors are the recently introduced ELKO \cite{Ahluwalia:2004ab,Ahluwalia:2004sz,Ahluwalia:2016jwz,Ahluwalia:2016rwl,daRocha:2007pz,HoffdaSilva:2009is,daRocha:2008we,Fabbri:2010ws} which seem promising candidates for a dark matter completion of the standard model albeit also these spinors have not been observed.

For spinorial fields in general, a fundamental tool used for their analysis is the polar form. A spinor field written in polar form is just a spinor in which each component is written as the product of a module times a unitary phase while preserving manifest covariance. Writing spinors in polar form has several advantages: first, they are written in terms of real covariant quantities only, regardless their frame or representation \cite{Fabbri:2016msm,Fabbri:2018crr}; second, it is rather easy to obtain more general types of solutions \cite{Fabbri:2016laz,Fabbri:2019kfr}; finally, it is more straightforward to examine properties like the non-relativistic or macroscopic approximations \cite{Fabbri:2020ypd}. The case of singular spinor fields has also been studied, with flag-dipoles investigated in \cite{Fabbri:2020elt}: specifically, flagpoles and dipoles are studied in \cite{Fabbri:2017xyk}. ELKO are addressed in \cite{Fabbri:2019vut}.

One advantage of the polar form is that in it it becomes clear what are the true degrees of freedom and what are the components that can always be transferred into the underlying frame. A first thing we will do in this paper is to show that these last components are by construction the Goldstone modes of the system. A second thing will be to show that Majorana and Weyl spinors have no true degree of freedom and being completely transferable into the frame they can be seen as pure Goldstone states.
\section{An Illuminating Example: the Standard Model}
In order to illustrate the principle and concept we shall need for the spinor fields, we begin by giving the example of the doublet of complex scalar fields undergoing to the $\mathrm{SU(2)\!\times\!U(1)}$ transformations since this is the prototypical case of the known standard model of particle physics.

In the most general form $\mathrm{SU(2)}$ transformations are
\begin{eqnarray}
\boldsymbol{U}\!=\!e^{-\frac{i}{2}\vec{\boldsymbol{\sigma}}\cdot\vec{\theta}}
\end{eqnarray}
and defining
\begin{eqnarray}
y\!=\!\left|\vec{\theta}/2\right|
\end{eqnarray}
and so
\begin{eqnarray}
X\!=\!\cos{y}\\
\vec{Z}\!=\!\frac{1}{2}\frac{\sin{y}}{y}\vec{\theta}
\end{eqnarray}
which verify 
\begin{eqnarray}
X^{2}\!+\!\vec{Z}\cdot\vec{Z}\!=\!1
\end{eqnarray}
we can write
\begin{eqnarray}
\boldsymbol{U}\!=\!X\mathbb{I}\!-\!i\vec{Z}\!\cdot\!\vec{\boldsymbol{\sigma}}
\end{eqnarray}
in explicit form. The inverse is given by
\begin{eqnarray}
\boldsymbol{U}^{-1}\!=\!e^{\frac{i}{2}\vec{\boldsymbol{\sigma}}\cdot\vec{\theta}}
\end{eqnarray}
and therefore
\begin{eqnarray}
\boldsymbol{U}^{-1}\!=\!X\mathbb{I}\!+\!i\vec{Z}\!\cdot\!\vec{\boldsymbol{\sigma}}
\end{eqnarray}
as identity $X^{2}\!+\!\vec{Z}\cdot\vec{Z}\!=\!1$ would show.

The most general form of $\mathrm{SU(2)\!\times\!U(1)}$ transformations are then given by
\begin{eqnarray}
\boldsymbol{S}\!=\!\boldsymbol{U}e^{\frac{i}{2}\alpha}
\!=\!(X\mathbb{I}\!-\!i\vec{Z}\!\cdot\!\vec{\boldsymbol{\sigma}})e^{\frac{i}{2}\alpha}
\end{eqnarray}
as obvious since the two transformations commute.

Notice that it is easy to check that
\begin{eqnarray}
(U)^{a}_{\phantom{a}b}\boldsymbol{S}\boldsymbol{\sigma}^{b}\boldsymbol{S}^{-1}\!=\!\boldsymbol{\sigma}^{a}\label{const}
\end{eqnarray}
where $(U)^{a}_{\phantom{a}b}$ such that $(U)^{a}_{\phantom{a}k}(U^{T})^{k}_{\phantom{k}b}\!=\!\delta^{a}_{b}$ is a transformation that belongs to the $\mathrm{SO(3)}$ group and that is the real representation of the transformation of the $\mathrm{SU(2)}$ group.

The doublet of complex scalar fields transforms as
\begin{eqnarray}
&\Phi\!\rightarrow\!\boldsymbol{S}\Phi
\end{eqnarray}
in the most general case.

It can be seen that by calling
\begin{eqnarray}
(\partial_{\mu}X\vec{Z}-X\partial_{\mu}\vec{Z})\!+\!\vec{Z}\!\times\!\partial_{\mu}\vec{Z}
\!=\!-\frac{1}{2}\partial_{\mu}\vec{\zeta}
\end{eqnarray}
we can write 
\begin{eqnarray}
\boldsymbol{U}^{-1}\partial_{\mu}\boldsymbol{U}\!=\!-\frac{i}{2}\partial_{\mu}\vec{\zeta}
\!\cdot\!\vec{\boldsymbol{\sigma}}\label{auxU}
\end{eqnarray}
having used $X^{2}\!+\!\vec{Z}\cdot\vec{Z}\!=\!1$ in the computations.

We can introduce two types of gauge fields $\vec{A}_{\mu}$ and $B_{\mu}$ defined as what transforms like
\begin{eqnarray}
&g\vec{A}_{\mu}\!\cdot\!\vec{\boldsymbol{\sigma}}\!\rightarrow\!\boldsymbol{U}
\left[(g\vec{A}_{\mu}\!-\!\partial_{\mu}\vec{\zeta})
\!\cdot\!\vec{\boldsymbol{\sigma}}\right]\boldsymbol{U}^{-1}\label{A}\\ 
&g'B_{\mu}\!\rightarrow\!g'B_{\mu}\!-\!\partial_{\mu}\alpha\label{B}
\end{eqnarray}
so that
\begin{eqnarray}
&D_{\mu}\Phi\!=\!\nabla_{\mu}\Phi\!-\!\frac{i}{2}\left(g\vec{A}_{\mu}
\!\cdot\!\vec{\boldsymbol{\sigma}}\!-\!g'B_{\mu}\mathbb{I}\right)\Phi
\end{eqnarray}
is the gauge covariant derivative of the doublet of complex scalar fields, as a direct computation would show.

The dynamics is given by the Lagrangian 
\begin{eqnarray} 
&\mathscr{L}\!=\!D_{\mu}\Phi^{\dagger}D^{\mu}\Phi
\!+\!\lambda^{2}\left(v^{2}\Phi^{2}\!-\!\frac{1}{2}\Phi^{4}\right)
\end{eqnarray}
written in terms of the $v^{2}$ and $\lambda^{2}$ constants.

Because of the gauge covariant character of the derivatives such a Lagrangian is fully $\mathrm{SU(2)\!\times\!U(1)}$ invariant.

Considering now the most general form with which the doublet of complex scalar fields can be written
\begin{eqnarray}
\Phi\!=\!\left(\begin{tabular}{c}
$ae^{i\alpha}$\\ 
$be^{i\beta}$
\end{tabular}\right)
\label{Higgs}
\end{eqnarray}
and the most general $\mathrm{SU(2)\!\times\!U(1)}$ transformations, one can take a complex rotation around the second axis with angle $\theta$ and a complex rotation around the third axis with angle $\varphi$ giving the resulting general transformation law
\begin{eqnarray}
&\!\!\!\!\Phi\!\rightarrow\!\left(\begin{array}{cccc}
\!\cos{\theta/2}\!&\!\sin{\theta/2}\\
\!-\sin{\theta/2}\!&\!\cos{\theta/2}
\end{array}\right)\!\!\left(\begin{array}{cccc}
\!e^{i\varphi/2}\!&\!0\\
\!0\!&\!e^{-i\varphi/2}
\end{array}\right)\!\!\left(\begin{tabular}{c}
$ae^{i\alpha}$\\ 
$be^{i\beta}$
\end{tabular}\right)
\end{eqnarray}
which shows that if we pick the angles
\begin{eqnarray}
&\cot{\theta/2}\!=\!a/b\\
&\varphi\!=\!\beta-\alpha
\end{eqnarray}
then we have the specific transformation
\begin{eqnarray}
&\Phi\!\rightarrow\!\sqrt{a^{2}+b^{2}}e^{i(\beta+\alpha)/2}\left(\!\begin{tabular}{c}
$1$\\
$0$
\end{tabular}\!\right)
\end{eqnarray}
as easy to see. With another rotation or an abelian transformation of angle $\varsigma\!=\!-(\beta\!+\!\alpha)$ we get
\begin{eqnarray}
&\Phi\!\rightarrow\!\sqrt{a^{2}+b^{2}}\left(\!\begin{tabular}{c}
$1$\\
$0$
\end{tabular}\!\right)
\end{eqnarray}
also quite clearly. The same could be done if we intended to keep the lower component. In any case, one can always find a gauge in which for example
\begin{eqnarray}
\Phi_{\mathrm{polar}}\!=\!\phi\left(\begin{tabular}{c}
$0$\\ 
$1$
\end{tabular}\right)
\end{eqnarray}
where $\phi$ is a general real scalar field. Because such form is obtained by employing only the $\mathrm{SU(2)\!\times\!U(1)}$ transformations then we have that we can write
\begin{eqnarray}
\Phi_{\mathrm{polar}}\!=\!\boldsymbol{R}\Phi
\end{eqnarray}
for some $\boldsymbol{R}$ depending on the components of the general doublet of complex scalar fields.\! In conclusion we can say that for the most general doublet of complex scalar fields transforming under $\mathrm{SU(2)\!\times\!U(1)}$ we can always find one gauge called unitary gauge in which
\begin{eqnarray}
\Phi\!=\!\phi\boldsymbol{R}^{-1}\left(\begin{tabular}{c}
$0$\\ 
$1$
\end{tabular}\right)
\end{eqnarray}
called polar form, for some $\boldsymbol{R}$ in the $\mathrm{SU(2)\!\times\!U(1)}$ group and in terms of $\phi$ being a generic real scalar field,\! and the only degree of freedom,\! called module. The unitary gauge is the special gauge where the doublet of complex scalar fields can be written in polar form, the form for which its $4$ real components are re-arranged into that very special configuration where the real scalar degree of freedom is isolated from the $3$ real components that are moved into the gauge.\!\! These $3$ real components are encoded as the $3$ parameters of the $\boldsymbol{R}$ matrix and so the Goldstone bosons.

Remark that on the polar form the action of $\boldsymbol{U}_{3}$ or the abelian transformation have the same effect. This redundancy is at the basis of the fact that we cannot remove all components from the doublet of complex scalar fields.

The components that can be transferred into the gauge are transferred into the gauge potentials given by
\begin{eqnarray}
\boldsymbol{R}^{-1}\partial_{\mu}\boldsymbol{R}
\!=\!-\frac{i}{2}\partial_{\mu}\vec{\xi}\!\cdot\!\vec{\boldsymbol{\sigma}}
\!+\!\frac{i}{2}\partial_{\mu}\xi\mathbb{I}\label{auxR}
\end{eqnarray}
where $\xi$ and $\vec{\xi}$ are the Goldstone modes. Then, defining
\begin{eqnarray}
&g\vec{M}_{\mu}\!=\!g\vec{A}_{\mu}\!-\!\partial_{\mu}\vec{\xi}\label{M}\\ 
&g'N_{\mu}\!=\!g'B_{\mu}\!-\!\partial_{\mu}\xi\label{N}
\end{eqnarray}
we have that the gauge covariant derivative is
\begin{eqnarray}
&D_{\mu}\Phi\!=\!\left[\nabla_{\mu}\ln{\phi}
\!-\!\frac{i}{2}\left(g\vec{M}_{\mu}\!\cdot\!\vec{\boldsymbol{\sigma}}
\!-\!g'N_{\mu}\mathbb{I}\right)\right]\Phi
\end{eqnarray}
in the most general case possible. It is now the moment to see what are the transformations for the $\vec{M}_{\mu}$ and $N_{\mu}$ vectorial fields. We start by seeing that for an $\mathrm{SU(2)\!\times\!U(1)}$ transformation the polar form transforms according to 
\begin{eqnarray}
\boldsymbol{R}^{-1}\Phi_{\mathrm{polar}}
\!\rightarrow\!\boldsymbol{S}\boldsymbol{R}^{-1}\Phi_{\mathrm{polar}}
\end{eqnarray}
which means that $\boldsymbol{R}^{-1}\!\rightarrow\!\boldsymbol{S}\boldsymbol{R}^{-1}$ since $\Phi_{\mathrm{polar}}$ can not have any transformation. This means that 
\begin{eqnarray}
&\boldsymbol{R}\!\rightarrow\!\boldsymbol{R}\boldsymbol{S}^{-1}
\end{eqnarray}
and thus
\begin{eqnarray}
&\boldsymbol{R}^{-1}\partial_{\mu}\boldsymbol{R}
\!\rightarrow\!\boldsymbol{S}\left(\boldsymbol{R}^{-1}\partial_{\mu}\boldsymbol{R}
\!+\!\partial_{\mu}\boldsymbol{S}^{-1}\boldsymbol{S}\right)\boldsymbol{S}^{-1}
\end{eqnarray}
as clear. Because $\boldsymbol{S}\!=\!\boldsymbol{U}e^{\frac{i}{2}\alpha}$ we have
\begin{eqnarray}
\partial_{\mu}\boldsymbol{S}^{-1}\boldsymbol{S}
\!=\!\partial_{\mu}\boldsymbol{U}^{-1}\boldsymbol{U}\!-\!\frac{i}{2}\partial_{\mu}\alpha
\end{eqnarray}
so that employing (\ref{auxU}) and (\ref{auxR}) we can write
\begin{eqnarray}
\nonumber
&-\partial_{\mu}\vec{\xi}\!\cdot\!\vec{\boldsymbol{\sigma}}
\!+\!\partial_{\mu}\xi\mathbb{I}\!\rightarrow\!
\boldsymbol{S}(-\partial_{\mu}\vec{\xi}\!\cdot\!\vec{\boldsymbol{\sigma}}
\!+\!\partial_{\mu}\xi\mathbb{I}+\\
&+\partial_{\mu}\vec{\zeta}\!\cdot\!\vec{\boldsymbol{\sigma}}
\!-\!\partial_{\mu}\alpha)\boldsymbol{S}^{-1}
\end{eqnarray}
after some simplification. With (\ref{A}-\ref{B}) we then get
\begin{eqnarray}
\nonumber
&g\vec{A}_{\mu}\!\cdot\!\vec{\boldsymbol{\sigma}}
\!-\!g'B_{\mu}\mathbb{I}
\!-\!\partial_{\mu}\vec{\xi}\!\cdot\!\vec{\boldsymbol{\sigma}}
\!+\!\partial_{\mu}\xi\mathbb{I}\rightarrow\\
\nonumber
&\rightarrow\boldsymbol{S}(g\vec{A}_{\mu}\!\cdot\!\vec{\boldsymbol{\sigma}}
\!-\!\partial_{\mu}\vec{\zeta}\!\cdot\!\vec{\boldsymbol{\sigma}}
\!-\!g'B_{\mu}\mathbb{I}\!+\!\partial_{\mu}\alpha\mathbb{I}-\\
\nonumber
&-\partial_{\mu}\vec{\xi}\!\cdot\!\vec{\boldsymbol{\sigma}}
\!+\!\partial_{\mu}\xi\mathbb{I}
\!+\!\partial_{\mu}\vec{\zeta}\!\cdot\!\vec{\boldsymbol{\sigma}}
\!-\!\partial_{\mu}\alpha\mathbb{I})\boldsymbol{S}^{-1}\equiv\\
&\equiv\boldsymbol{S}(g\vec{A}_{\mu}\!\cdot\!\vec{\boldsymbol{\sigma}}
\!-\!g'B_{\mu}\mathbb{I}
\!-\!\partial_{\mu}\vec{\xi}\!\cdot\!\vec{\boldsymbol{\sigma}}
\!+\!\partial_{\mu}\xi\mathbb{I})\boldsymbol{S}^{-1}
\end{eqnarray}
identically. In terms of definitions (\ref{M}-\ref{N}) we can express
\begin{eqnarray}
&g\vec{M}_{\mu}\!\cdot\!\vec{\boldsymbol{\sigma}}\!-\!g'N_{\mu}\mathbb{I}\!\rightarrow\!
\boldsymbol{S}(g\vec{M}_{\mu}\!\cdot\!\vec{\boldsymbol{\sigma}}\!-\!g'N_{\mu}\mathbb{I})\boldsymbol{S}^{-1}
\end{eqnarray}
then due to the linear independence of identity and Pauli matrices we can split
\begin{eqnarray}
&\vec{M}_{\mu}\!\cdot\!\vec{\boldsymbol{\sigma}}\!\rightarrow\!
\boldsymbol{S}\vec{M}_{\mu}\!\cdot\!\vec{\boldsymbol{\sigma}}\boldsymbol{S}^{-1}\\
&N_{\mu}\!\rightarrow\!N_{\mu}
\end{eqnarray}
and finally
\begin{eqnarray}
&M_{\mu}^{a}\!\rightarrow\!(U)^{a}_{\phantom{a}b}M_{\mu}^{b}\\
&N_{\mu}\!\rightarrow\!N_{\mu}
\end{eqnarray}
showing that $M_{\mu}^{a}$ and $N_{\mu}$ no longer display the transformation of gauge fields and they now transform as vector fields. As we said above, whenever we write the doublet of complex scalar fields in polar form we can keep the real scalar degree of freedom isolated from the $3$ real components that can be transferred into the gauge fields. After that this is done we say that the $3$ Goldstone bosons are absorbed into $3$ gauge fields becoming the $3$ longitudinal components of what are now $3$ vector fields, each with $3$ components, and which no longer transform as the gauge fields but simply as vectorial fields. Such a process is just the Goldstone mechanism, or the Goldstone theorem.

Notice that the aforementioned redundancy for the allowed transformations will leave an additional symmetry permitted for the doublet of complex scalar fields.

Let us move to the dynamics. Writing the Lagrangian in terms of the polar form gives
\begin{eqnarray} 
\nonumber
&\mathscr{L}\!=\!\nabla_{\mu}\phi\nabla^{\mu}\phi
\!+\!\frac{1}{4}\phi^{2}(g^{2}\vec{M}_{\mu}\!\cdot\!\vec{M}^{\mu}\!+\!g'^{2}N^{\mu}N_{\mu}-\\
&-2gg'N^{\mu}\vec{M}_{\mu}\!\cdot\!\Phi^{\dagger}\vec{\boldsymbol{\sigma}}\Phi/\phi^{2})
\!+\!\lambda^{2}\left(v^{2}\phi^{2}\!-\!\frac{1}{2}\phi^{4}\right)
\end{eqnarray}
in terms of the normalized isospin vector $\Phi^{\dagger}\vec{\boldsymbol{\sigma}}\Phi/\phi^{2}$ which must now be computed. In unitary gauge it is 
\begin{eqnarray}
&\Phi^{\dagger}\boldsymbol{\sigma}^{a}\Phi/\phi^{2}
\!=\!(0\ 1)\boldsymbol{\sigma}^{a}\left(\begin{tabular}{c}
$0$\\ 
$1$
\end{tabular}\right)\!=\!-\left(\begin{tabular}{c}
$0$\\ 
$0$\\
$1$
\end{tabular}\right)
\end{eqnarray}
and thus
\begin{eqnarray}
&-2gg'N_{\mu}M^{\mu}_{a}\Phi^{\dagger}\boldsymbol{\sigma}^{a}\Phi/\phi^{2}
\!=\!2gg'N_{\mu}M^{\mu}_{3}\label{mix}
\end{eqnarray}
but because the left-hand side of this last expression is a gauge scalar then (\ref{mix}) is valid in general. By plugging it into the Lagrangian we get
\begin{eqnarray} 
\nonumber
&\mathscr{L}\!=\!\nabla_{\mu}\phi\nabla^{\mu}\phi
\!+\!\frac{1}{4}\phi^{2}[g^{2}(M_{\mu}^{1}M^{\mu}_{1}\!+\!M_{\mu}^{2}M^{\mu}_{2})+\\
&+(gM_{\mu}^{3}\!+\!g'N_{\mu})(gM^{\mu}_{3}\!+\!g'N^{\mu})]
\!+\!\lambda^{2}\!\left(v^{2}\phi^{2}\!-\!\frac{1}{2}\phi^{4}\right)
\end{eqnarray}
in which by re-naming the vector fields as
\begin{eqnarray}
&\frac{1}{\sqrt{2}}\left(M^{1}_{\mu}\pm iM^{2}_{\mu}\right)=W_{\mu}^{\pm}
\end{eqnarray}
and
\begin{eqnarray}
&gM^{\mu}_{3}\!+\!g'N^{\mu}=\sqrt{g^{2}\!+\!g'^{2}}Z^{\mu}\\
&-g'M^{\mu}_{3}\!+\!gN^{\mu}=\sqrt{g^{2}\!+\!g'^{2}}A^{\mu}
\end{eqnarray}
we eventually obtain
\begin{eqnarray} 
\nonumber
&\mathscr{L}\!=\!\nabla_{\mu}\phi\nabla^{\mu}\phi
\!+\!\frac{1}{4}\phi^{2}[2g^{2}W^{+}W^{-}\!+\!(g^{2}\!+\!g'^{2})Z^{2}]+\\
&+\lambda^{2}\!\left(v^{2}\phi^{2}\!-\!\frac{1}{2}\phi^{4}\right)
\end{eqnarray}
in which $3$ vector fields have acquired quadratic terms.

The above-mentioned additional symmetry is the one corresponding to the gauge field with no quadratic term.

To finally implement spontaneous breaking of the symmetry, we have to assume the shift
\begin{eqnarray}
&\phi\!=\!v\!+\!H
\end{eqnarray}
and substitute this in the Lagrangian. In the terms that are quadratic in the vector fields we can find the terms
\begin{eqnarray}
&\mathscr{L}_{\mathrm{mass}}\!=\!m_{W}^{2}W^{+}W^{-}\!+\!\frac{1}{2}m_{Z}^{2}Z^{2}
\end{eqnarray}
where $gv\!=\!m_{W}\sqrt{2}$ and $v\sqrt{g^{2}\!+\!g'^{2}}\!=\!m_{Z}\sqrt{2}$ as are given in the standard model of particle physics. So eventually, the spontaneous breaking of the symmetry is the mechanism that generates the masses of all the gauge fields.

So the Goldstone mechanism is the procedure for which the doublet of complex scalar fields is re-arranged in such a way as to retain only its real scalar degree of freedom while the remaining three components are the Goldstone modes that can be transferred into the gauge fields as the longitudinal components. Then the Higgs mechanism is the procedure with which the breakdown of the symmetry fixes the ground state giving mass to the gauge fields.

This is just the standard model of particle physics. 

We have only presented it in a way that will help us in highlighting interesting properties for spinorial fields.
\section{The General Spinorial Fields}
Having this example at hand, we will now move toward the main section of the work by studying the spinor fields, defined as the quadruplet of complex scalar fields transforming under the complex Lorentz transformations.

In the most general form complex Lorentz transformations are given in terms of Clifford matrices $\boldsymbol{\gamma}_{a}$ such that 
\begin{eqnarray}
&\{\boldsymbol{\gamma}_{a},\!\boldsymbol{\gamma}_{b}\}\!=\!2\mathbb{I}\eta_{ab}
\end{eqnarray}
where $\eta_{ab}$ is the Minkowski matrix. Then we can define
\begin{eqnarray}
&\frac{1}{4}\left[\boldsymbol{\gamma}_{a},\!\boldsymbol{\gamma}_{b}\right]
\!=\!\boldsymbol{\sigma}_{ab}
\end{eqnarray}
where $\boldsymbol{\sigma}_{ab}$ also verify
\begin{eqnarray}
&2i\boldsymbol{\sigma}_{ab}\!=\!\varepsilon_{abcd}\boldsymbol{\pi}\boldsymbol{\sigma}^{cd}
\end{eqnarray}
implicitly defining the $\boldsymbol{\pi}$ matrix (this matrix is normally denoted as a gamma with an index five, but because in the space-time this index has no meaning we will employ a notation with no index). It possible to see that
\begin{eqnarray}
&\boldsymbol{\gamma}_{i}\boldsymbol{\gamma}_{j}\boldsymbol{\gamma}_{k}
\!=\!\boldsymbol{\gamma}_{i}\eta_{jk}-\boldsymbol{\gamma}_{j}\eta_{ik}
\!+\!\boldsymbol{\gamma}_{k}\eta_{ij}
\!+\!i\varepsilon_{ijkq}\boldsymbol{\pi}\boldsymbol{\gamma}^{q}
\end{eqnarray}
from which it is possible to prove that
\begin{eqnarray}
&\{\boldsymbol{\gamma}_{a},\boldsymbol{\sigma}_{bc}\}
=i\varepsilon_{abcd}\boldsymbol{\pi}\boldsymbol{\gamma}^{d}\label{anticommgamma}\\
&[\boldsymbol{\gamma}_{a},\boldsymbol{\sigma}_{bc}]
=\eta_{ab}\boldsymbol{\gamma}_{c}\!-\!\eta_{ac}\boldsymbol{\gamma}_{b}\label{commgamma}
\end{eqnarray}
and
\begin{eqnarray}
&\{\boldsymbol{\sigma}_{ab},\boldsymbol{\sigma}_{cd}\}
=\frac{1}{2}[(\eta_{ad}\eta_{bc}\!-\!\eta_{ac}\eta_{bd})\mathbb{I}
\!+\!i\varepsilon_{abcd}\boldsymbol{\pi}]\label{anticommsigma}\\
&[\boldsymbol{\sigma}_{ab},\boldsymbol{\sigma}_{cd}]
=\eta_{ad}\boldsymbol{\sigma}_{bc}\!-\!\eta_{ac}\boldsymbol{\sigma}_{bd}
\!+\!\eta_{bc}\boldsymbol{\sigma}_{ad}\!-\!\eta_{bd}\boldsymbol{\sigma}_{ac}\label{commsigma}
\end{eqnarray}
are all valid as geometric identities.

This very last relationship in particular tells us that the $\boldsymbol{\sigma}_{ab}$ matrices are the generators of the Lorentz algebra, so that with parameters $\theta_{ij}\!=\!-\theta_{ji}$ we can write 
\begin{eqnarray}
&\boldsymbol{\Lambda}\!=\!e^{\frac{1}{2}\theta_{ab}\boldsymbol{\sigma}^{ab}}
\end{eqnarray}
as the Lorentz transformations in the most general case and which can be made explicit. For that we define
\begin{eqnarray}
a\!=\!-\frac{1}{8}\theta_{ij}\theta^{ij}\\
b\!=\!\frac{1}{16}\theta_{ij}\theta_{ab}\varepsilon^{ijab}
\end{eqnarray}
and then
\begin{eqnarray}
2x^{2}\!=\!a\!+\!\sqrt{a^{2}\!+\!b^{2}}\\
2y^{2}\!=\!-a\!+\!\sqrt{a^{2}\!+\!b^{2}}
\end{eqnarray}
so to introduce
\begin{eqnarray}
\cos{y}\cosh{x}\!=\!X\\
\sin{y}\sinh{x}\!=\!Y\\
\nonumber
\left(\frac{x\sinh{x}\cos{y}
+y\sin{y}\cosh{x}}{x^{2}+y^{2}}\right)\theta^{ab}+\\
+\left(\frac{x\cosh{x}\sin{y}
-y\cos{y}\sinh{x}}{x^{2}+y^{2}}\right)\!\frac{1}{2}\theta_{ij}\varepsilon^{ijab}\!=\!Z^{ab}
\end{eqnarray}
which verify
\begin{eqnarray}
X^{2}\!-\!Y^{2}\!+\!\frac{1}{8}Z^{ab}Z_{ab}\!=\!1\\
2XY\!-\!\frac{1}{16}Z^{ij}Z^{ab}\varepsilon_{ijab}\!=\!0
\end{eqnarray}
in terms of which using (\ref{anticommsigma}) we can see that
\begin{eqnarray}
\boldsymbol{\Lambda}\!=\!
X\mathbb{I}\!+\!Yi\boldsymbol{\pi}+\frac{1}{2}Z^{ab}\boldsymbol{\sigma}_{ab}
\end{eqnarray}
in the most compact way. The inverse is
\begin{eqnarray}
&\boldsymbol{\Lambda}^{-1}\!=\!e^{-\frac{1}{2}\theta_{ab}\boldsymbol{\sigma}^{ab}}
\end{eqnarray}
written explicitly as
\begin{eqnarray}
\boldsymbol{\Lambda}^{-1}\!=\!X\mathbb{I}\!+\!Yi\boldsymbol{\pi}-\frac{1}{2}Z^{ab}\boldsymbol{\sigma}_{ab}
\end{eqnarray}
as clear after using relations $8X^{2}\!-\!8Y^{2}\!+\!Z^{ab}Z_{ab}\!=\!8$ and $32XY\!-\!Z^{ij}Z^{ab}\varepsilon_{ijab}\!=\!0$ as they are given here above.

The complete Lorentz and phase transformation is
\begin{eqnarray}
&\boldsymbol{S}\!=\!\boldsymbol{\Lambda}e^{iq\alpha}\!=\!
(X\mathbb{I}\!+\!Yi\boldsymbol{\pi}\!+\!\frac{1}{2}Z^{ab}\boldsymbol{\sigma}_{ab})e^{iq\alpha}
\end{eqnarray}
and it is called spinorial transformation.

Notice that we have
\begin{eqnarray}
(\Lambda)^{a}_{\phantom{a}b}\boldsymbol{S}\boldsymbol{\gamma}^{b}\boldsymbol{S}^{-1}\!=\!\boldsymbol{\gamma}^{a}\label{constgamma}
\end{eqnarray}
where $(\Lambda)^{a}_{\phantom{a}b}$ such that $(\Lambda)^{a}_{\phantom{a}k}(\Lambda)^{b}_{\phantom{b}j}\eta^{kj}\!=\!\eta^{ab}$ is a transformation that belongs to the $\mathrm{SO(1,3)}$ group and that is the real representation of the Lorentz transformation.

With this transformation we can define spinor fields as what transforms according to 
\begin{eqnarray}
&\psi\!\rightarrow\!\boldsymbol{S}\psi
\end{eqnarray}
in the most general case.

By introducing the object
\begin{eqnarray}
\nonumber
&(\partial_{\mu}XZ^{ab}-X\partial_{\mu}Z^{ab})
+\frac{1}{2}(\partial_{\mu}YZ_{ij}-Y\partial_{\mu}Z_{ij})\varepsilon^{ijab}+\\
&+\partial_{\mu}Z^{ak}Z^{b}_{\phantom{b}k}\!=\!-\partial_{\mu}\zeta^{ab}
\end{eqnarray}
we can finally write
\begin{eqnarray}
\boldsymbol{\Lambda}^{-1}\partial_{\mu}\boldsymbol{\Lambda}
\!=\!\frac{1}{2}\partial_{\mu}\zeta_{ab}\boldsymbol{\sigma}^{ab}
\end{eqnarray}
with $8X^{2}\!-\!8Y^{2}\!+\!Z^{ab}Z_{ab}\!=\!8$ and $32XY\!-\!Z^{ij}Z^{ab}\varepsilon_{ijab}\!=\!0$ which have been used throughout the whole calculation.

We can now introduce the spinor connection $\Omega_{ij\mu}$ given in terms of its transformation 
\begin{eqnarray}
&\frac{1}{2}\Omega_{ij\mu}\boldsymbol{\sigma}^{ij}\!\rightarrow\!\boldsymbol{\Lambda}
\left[\frac{1}{2}(\Omega_{ij\mu}\!-\!\partial_{\mu}\zeta_{ij})\boldsymbol{\sigma}^{ij}\right]\boldsymbol{\Lambda}^{-1}\\
&A_{\mu}\!\rightarrow\!A_{\mu}\!-\!\partial_{\mu}\alpha
\end{eqnarray}
so that
\begin{eqnarray}
&\boldsymbol{\nabla}_{\mu}\psi\!=\!\partial_{\mu}\psi\!+\!\frac{1}{2}\Omega_{ij\mu}\boldsymbol{\sigma}^{ij}\psi
\!+\!iqA_{\mu}\psi
\end{eqnarray}
is the spinorial covariant derivative of the spinor fields.

The Dirac matter field equations are
\begin{eqnarray}
&i\boldsymbol{\gamma}^{\mu}\boldsymbol{\nabla}_{\mu}\psi
\!-\!XW_{\mu}\boldsymbol{\gamma}^{\mu}\boldsymbol{\pi}\psi\!-\!m\psi\!=\!0
\label{D}
\end{eqnarray}
with $W_{\mu}$ axial-vector torsion and $X$ torsion-spin coupling constant, added to be in the most general case.
\subsection{Regular Spinor Fields}
It is time to find the unitary frame in which the spinor can be written in polar form. The first thing to do is to observe that in general the two bi-linear spinor quantities $i\overline{\psi}\boldsymbol{\pi}\psi$ and $\overline{\psi}\psi$ are not both equal to zero identically, and therefore we can always find a frame in which
\begin{eqnarray}
&\!\psi\!=\!\phi e^{-\frac{i}{2}\beta\boldsymbol{\pi}}
\boldsymbol{L}^{-1}\left(\!\begin{tabular}{c}
$1$\\
$0$\\
$1$\\
$0$
\end{tabular}\!\right)
\label{spinorch}
\end{eqnarray}
so that
\begin{eqnarray}
&i\overline{\psi}\boldsymbol{\pi}\psi\!=\!2\phi^{2}\sin{\beta}\\
&\overline{\psi}\psi\!=\!2\phi^{2}\cos{\beta}
\end{eqnarray}
as well as 
\begin{eqnarray}
&\overline{\psi}\boldsymbol{\gamma}^{a}\boldsymbol{\pi}\psi\!=\!2\phi^{2}s^{a}\\
&\overline{\psi}\boldsymbol{\gamma}^{a}\psi\!=\!2\phi^{2}u^{a}
\end{eqnarray}
with $\boldsymbol{L}$ Lorentz transformation while $\phi$ and $\beta$ are a scalar and a pseudo-scalar and the only true degrees of freedom called module and Yvon-Takabayashi angle. So the $8$ real components of the spinor can be  rearranged in such a way that the $2$ real scalar degrees of freedom are isolated from the $6$ real components that can always be transferred into the frame. In the details, these $6$ real components are the $3$ rapidities and the $3$ Euler angles encoded as parameters of the Lorentz transformation $\boldsymbol{L}$ and therefore they could be identified as the Goldstone bosons for the spinor field, in the same way as the $3$ parameters of the $\boldsymbol{R}$ matrix are defined as the Goldstone bosons of the Higgs field above.

The components that will be transferred into the frame are transferred into the spinorial connection
\begin{eqnarray}
\boldsymbol{L}^{-1}\partial_{\mu}\boldsymbol{L}\!=\!iq\partial_{\mu}\xi\mathbb{I}
\!+\!\frac{1}{2}\partial_{\mu}\xi^{ab}\boldsymbol{\sigma}_{ab}\label{LdL}
\end{eqnarray}
where $\xi$ and $\xi^{ab}$ are the exact mathematical analogous of the Goldstone modes defined in general. Then, setting
\begin{eqnarray}
&q(\partial_{\mu}\xi\!-\!A_{\mu})\!\equiv\!P_{\mu}\label{P}\\
&\partial_{\mu}\xi_{ij}\!-\!\Omega_{ij\mu}\!\equiv\!R_{ij\mu}\label{R}
\end{eqnarray}
we have that the spinorial covariant derivative is
\begin{eqnarray}
&\!\!\!\!\!\!\!\!\boldsymbol{\nabla}_{\mu}\psi\!=\!(-\frac{i}{2}\nabla_{\mu}\beta\boldsymbol{\pi}
\!+\!\nabla_{\mu}\ln{\phi}\mathbb{I}
\!-\!iP_{\mu}\mathbb{I}\!-\!\frac{1}{2}R_{ij\mu}\boldsymbol{\sigma}^{ij})\psi
\label{decspinder}
\end{eqnarray}
in the most general case. We see that it does contain the derivatives of the two degrees of freedom given by module and Yvon-Takabayashi angle while the components that can be transferred into the frame combine with the spin connection and the gauge field.\! After that, the Goldstone bosons absorbed into the spin connection and the gauge field become the longitudinal components of the $P_{\mu}$ and $R_{ji\mu}$ objects, which no longer transform as one spin connection and one gauge field but simply as gauge invariant real tensor quantities. Again this is the Goldstone mechanism, and it can be proven for the spinor fields in exactly the same way as we did for the Higgs field.

Plugging the polar in the Dirac equations we obtain
\begin{eqnarray}
&\!\!\!\!B_{\mu}\!-\!2P^{\iota}u_{[\iota}s_{\mu]}\!+\!(\nabla\beta\!-\!2XW)_{\mu}
\!+\!2s_{\mu}m\cos{\beta}\!=\!0\label{dep1}\\
&\!\!\!\!R_{\mu}\!-\!2P^{\rho}u^{\nu}s^{\alpha}\varepsilon_{\mu\rho\nu\alpha}\!+\!2s_{\mu}m\sin{\beta}
\!+\!\nabla_{\mu}\ln{\phi^{2}}\!=\!0\label{dep2}
\end{eqnarray}
with $R_{\mu a}^{\phantom{\mu a}a}\!=\!R_{\mu}$ and $\frac{1}{2}\varepsilon_{\mu\alpha\nu\iota}R^{\alpha\nu\iota}\!=\!B_{\mu}$ and which can be proven to be equivalent to the original Dirac equations in general. The Dirac equations are $8$ real equations and thus as many as the $2$ vectorial equations (\ref{dep1}, \ref{dep2}). Such a pair of vector equations specify all space-time derivatives for both the module and the Yvon-Takabayashi angle.

The explicit structure of the polar form of the spinorial field (\ref{spinorch}), the structure of its spinorial covariant derivative (\ref{decspinder}) and the field equations (\ref{dep1}, \ref{dep2}) have been thoroughly discussed in work \cite{Fabbri:2020ypd} and references therein.

So what is the physical information that we can extract from this section? The main idea is that in exact analogy with the case of the Higgs field also for spinor fields it is possible to write them in a way in which their degrees of freedom (the observable Higgs $H$ of the standard model and module and Yvon-Takabayashi angle $\phi$ and $\beta$ for the spinor field) are isolated inside the fields themselves while all the remaining components (the parameters of $\boldsymbol{R}$ in the standard model and of $\boldsymbol{L}$ for the spinor field) are moved into the gauge where they combine with the underlying connection to become real tensor fields (the $M_{\mu}^{a}$ and $N_{\mu}$ of the standard model and the $P_{\mu}$ and $R_{ji\mu}$ for the spinor fields). Either way, the transferred components (and that is the parameters of $\boldsymbol{R}$ in the standard model and of $\boldsymbol{L}$ for the spinor field) are identified with the Goldstone bosons.

If the dynamical action were to be subject to a mechanism of symmetry breaking, the tensor fields (the $M_{\mu}^{a}$ and $N_{\mu}$ of the standard model and the $P_{\mu}$ and $R_{ji\mu}$ for the spinor fields) would give rise to massive bosons, although the mechanism of symmetry breaking is not a necessity.

So we showed that the derivatives of the parameters of the local transformations that bring a field into its polar form are the Goldstone bosons of the system, whether we are in the known standard model or for spinor fields.

We next move on to study singular spinor fields.
\subsection{Singular Spinor Fields}
We now move on to study spinors in singular case, that is when both $i\overline{\psi}\boldsymbol{\pi}\psi\!=\!0$ and $\overline{\psi}\psi\!=\!0$ identically.

In this special instance, the polar decomposition is still possible. It gives that we can always write the spinor as
\begin{eqnarray}
&\psi\!=\!\frac{1}{\sqrt{2}}(\mathbb{I}\cos{\frac{\alpha}{2}}
\!-\!\boldsymbol{\pi}\sin{\frac{\alpha}{2}})\boldsymbol{L}^{-1}\left(\!\begin{tabular}{c}
$1$\\
$0$\\
$0$\\
$1$
\end{tabular}\!\right)
\label{singular}
\end{eqnarray}
so that
\begin{eqnarray}
&\overline{\psi}\boldsymbol{\gamma}^{k}\boldsymbol{\pi}\psi
\!=\!-\sin{\alpha}\overline{\psi}\boldsymbol{\gamma}^{k}\psi
\end{eqnarray}
and
\begin{eqnarray}
&\overline{\psi}\boldsymbol{\gamma}^{k}\psi\!=\!U^{k}\\
&2i\overline{\psi}\boldsymbol{\sigma}^{ij}\psi\!=\!M^{ij}
\end{eqnarray}
where $\boldsymbol{L}$ is a general Lorentz transformation and $\alpha$ a real pseudo-scalar and the only degree of freedom. Therefore, the two constraints $i\overline{\psi}\boldsymbol{\pi}\psi\!=\!\overline{\psi}\psi\!=\!0$ reduce $8$ components to $6$ only, but the presence of one degree of freedom is still ensured by a redundancy in the Lorentz transformations.

In general, still with (\ref{LdL}) and (\ref{P}-\ref{R}), we have
\begin{eqnarray}
\nonumber
&\boldsymbol{\nabla}_{\mu}\psi\!=\![-\frac{1}{2}(\mathbb{I}\tan{\alpha}
\!+\!\boldsymbol{\pi}\sec{\alpha})\nabla_{\mu}\alpha-\\
&-iP_{\mu}\mathbb{I}
\!-\!\frac{1}{2}R_{ij\mu}\boldsymbol{\sigma}^{ij}]\psi
\end{eqnarray}
as the spinorial covariant derivative.

The Dirac equations decompose as
\begin{eqnarray}
\nonumber
&\![(2XW\!-\!B)^{\sigma}\varepsilon_{\sigma\mu\rho\nu}\!+\!R_{[\mu}g_{\rho]\nu}+\\
&+g_{\nu[\mu}\nabla_{\rho]}\alpha\tan{\alpha}]
M_{\eta\zeta}\varepsilon^{\mu\rho\eta\zeta}\!=\!0\label{1}\\
\nonumber
&\!\![(2XW\!-\!B)^{\sigma}\varepsilon_{\sigma\mu\rho\nu}\!+\!R_{[\mu}g_{\rho]\nu}+\\
&+g_{\nu[\mu}\nabla_{\rho]}\alpha\tan{\alpha}]M^{\mu\rho}\!+\!4mU_{\nu}\!=\!0\label{2}\\
&\!\!\!\!(\varepsilon^{\mu\rho\sigma\nu}\nabla_{\mu}\alpha\sec{\alpha}
\!-\!2P^{[\rho}g^{\sigma]\nu})M_{\rho\sigma}\!=\!0\label{3}\\
&\!\!\!\!\!\!\!\!M_{\rho\sigma}(g^{\nu[\rho}\nabla^{\sigma]}\alpha\sec{\alpha}
\!-\!2P_{\mu}\varepsilon^{\mu\rho\sigma\nu})\!+\!4m\sin{\alpha}U^{\nu}\!=\!0\label{4}
\end{eqnarray}
specifying all derivatives of the degree of freedom. Notice that the above-mentioned redundancy in all the possible Lorentz transformation is here reflected as a redundancy in the number of all the independent field equations.

\subsubsection{Flag-dipole spinors}
The case of flag-dipoles is the first sub-class of interest, characterized by the fact that $\alpha$ is not constrained to have any specific value. Consequently, there still is one degree of freedom, and the Dirac equations specify its derivative.

These results can be found in reference \cite{Fabbri:2020elt}. One more time, what is the physical meaning of this? The answer is that with respect to regular spinor fields, singular spinor fields are restricted to have fewer degrees of freedom but still they have some, so they are still physical. As for the components transferred into the gauge and eaten by the connection, they formally are still the Goldstone bosons.

\subsubsection{Weyl spinors}
The case of dipoles is the second sub-class of interest, and in particular for us it is the first case displaying some peculiar behaviour. In fact, Weyl spinors are given when $\alpha\!=\!\pm\pi/2$ identically. Its form is therefore given by
\begin{eqnarray}
&\psi\!=\!\frac{1}{2}(\mathbb{I}\mp\boldsymbol{\pi})\boldsymbol{L}^{-1}\left(\!\begin{tabular}{c}
$1$\\
$0$\\
$0$\\
$1$
\end{tabular}\!\right)
\label{Weyl}
\end{eqnarray}
for left-handed and right-handed chiral parts. Let us now check this statement directly, considering the Weyl spinor written in general as
\begin{eqnarray}
&\psi_{L}\!=\!\left(\begin{tabular}{c}
$ae^{i\alpha}$\\ 
$be^{i\beta}$\\
$0$\\
$0$
\end{tabular}\right)
\end{eqnarray}
performing Lorentz transformations on it.\! Analogously to what we have done in the case of the Higgs field (\ref{Higgs}), the Weyl spinor is also, mathematically, a doublet of complex scalar functions, so that we can employ rotations in order to bring it in the form
\begin{eqnarray}
&\psi\!\rightarrow\!\phi\left(\!\begin{tabular}{c}
$1$\\
$0$\\
$0$\\
$0$
\end{tabular}\!\right)
\end{eqnarray}
with $\phi$ a general real function, exactly as we have had for the Higgs field. But differently from what we had for the Higgs field, for which we can only use complex rotations, for Weyl spinors we can also use complex boosts. Using a boost along the third axis with rapidity $\varphi\!=\!\ln{\phi^{2}}$ brings the Weyl spinor in the final form given by
\begin{eqnarray}
&\psi\!\rightarrow\!\left(\!\begin{tabular}{c}
$1$\\
$0$\\
$0$\\
$0$
\end{tabular}\!\right)
\end{eqnarray}
in general. For right-handed Weyl spinors we would have
\begin{eqnarray}
&\psi\!\rightarrow\!\left(\!\begin{tabular}{c}
$0$\\
$0$\\
$0$\\
$1$
\end{tabular}\!\right)
\end{eqnarray}
also in general. Thus for a Weyl spinor
\begin{eqnarray}
&\psi\!\rightarrow\!\frac{1}{2}(\mathbb{I}\mp\boldsymbol{\pi})\left(\!\begin{tabular}{c}
$1$\\
$0$\\
$0$\\
$1$
\end{tabular}\!\right)
\end{eqnarray}
in the most general case. So we can always write a Weyl spinor in the form (\ref{Weyl}). This shows that the Weyl spinor does not possess any degree of freedom whatsoever.

The spinorial covariant derivative is given by
\begin{eqnarray}
&\boldsymbol{\nabla}_{\mu}\psi
\!=\!(-iP_{\mu}\mathbb{I}\!-\!\frac{1}{2}R_{ij\mu}\boldsymbol{\sigma}^{ij})\psi
\end{eqnarray}
as straightforward to see. Of course no degree of freedom appears in the explicit form of the spinorial derivative.

The dynamical equations in this case impose $m\!=\!0$ and they can be written as
\begin{eqnarray}
&R_{\mu}U^{\mu}\!=\!0\\
&(-B_{\mu}\!+\!2XW_{\mu}\!\pm\!2P_{\mu})U^{\mu}\!=\!0\\
&[(-B_{\mu}\!+\!2XW_{\mu}\!\pm\!2P_{\mu})\varepsilon^{\mu\rho\alpha\nu}
\!+\!g^{\rho[\alpha}R^{\nu]}]U_{\rho}\!=\!0
\end{eqnarray}
which are constraining relationships on the Weyl spinor.

Therefore, differently from the case of regular spinorial fields, and also differently from singular spinor fields in a strict sense like the flag-dipoles, dipoles are restricted so much that they have no degree of freedom, and this is a situation that raises a few questions about their physical status. Weyl spinors are constituted only by components that can be transferred into the frame and swallowed by the connection, so that they are pure Goldstone states.

\subsubsection{Majorana spinors}
The case of flagpoles is the third sub-class of interest, the second displaying the peculiar behaviour above. The Majorana spinor has $\alpha\!=\!0$ identically. Its form is hence
\begin{eqnarray}
&\psi\!=\!\frac{1}{\sqrt{2}}\boldsymbol{L}^{-1}\left(\!\begin{tabular}{c}
$1$\\
$0$\\
$0$\\
$1$
\end{tabular}\!\right)
\end{eqnarray}
so that $i\boldsymbol{\gamma}^{2}\psi^{\ast}\!=\!\psi$ as eigen-spinor of charge-conjugation operator. The actual computations are analogous to the ones we did for the case of Weyl. And as for Weyl, there is no degree of freedom remaining in the spinor as well.

Because $i\boldsymbol{\gamma}^{2}\psi^{\ast}\!=\!\psi$ the spinorial covariant derivative is
\begin{eqnarray}
&\boldsymbol{\nabla}_{\mu}\psi\!=\!-\frac{1}{2}R_{ij\mu}\boldsymbol{\sigma}^{ij}\psi
\end{eqnarray}
having lost all contributions of the momentum too.

Because the Majorana spinor field has no spin then it decouples from torsion and we can assume torsion to be zero, so that the dynamical equations are
\begin{eqnarray}
&(g_{\sigma[\pi}B_{\kappa]}\!-\!R^{\mu}\varepsilon_{\mu\sigma\pi\kappa})M^{\pi\kappa}\!=\!0\\
&\frac{1}{2}(B_{\mu}\varepsilon^{\mu\sigma\pi\kappa}
\!+\!g^{\sigma[\pi}R^{\kappa]})M_{\pi\kappa}\!-\!2mU^{\sigma}\!=\!0
\end{eqnarray}
or equivalently
\begin{eqnarray}
&R_{\mu}U^{\mu}\!=\!0\\
&B_{\mu}U^{\mu}\!=\!0\\
&(-B_{\mu}\varepsilon^{\mu\rho\alpha\nu}
\!+\!g^{\rho[\alpha}R^{\nu]})U_{\rho}\!+\!2m\!M^{\alpha\nu}\!\!=\!0
\end{eqnarray}
which are analogous to what we had for the Weyl case as constraining relations on the Majorana spinor.

Therefore, differently from the case of regular spinorial fields, and also differently from singular spinor fields in a strict sense like the flag-dipoles, but similarly to dipoles, flagpoles are restricted to have no degree of freedom, with dire consequences for their physical status. Like the Weyl spinors, Majorana spinors have only components that can be transferred into the frame and swallowed by the connection, consequently being pure Goldstone states \cite{Fabbri:2017xyk}.

It is worth to mention, however, that, differently from dipoles, flagpoles may actually have a way out of such a circumstance. Indeed, special types of Majorana spinors, recently introduced and named ELKO, are defined with a different type of spinor adjoint that is independent from the original spinor \cite{Ahluwalia:2016rwl}. The consequence is that whereas Majorana spinors in polar form also have adjoint in polar form, ELKO spinors in polar form do not have adjoint in the polar form. This means that even in unitary gauge, where both a Majorana spinor and its adjoint would have no degree of freedom, either an ELKO or its adjoint will always have at least \emph{some} degree of freedom \cite{Fabbri:2019vut}.
\section{Conclusion}
In this paper, we have considered the standard model, reviewing the mechanism for which it is always possible to go in the unitary gauge, having all the non-physical components of the Higgs transferred into the vector bosons as their longitudinal degrees of freedom. Hence, we have seen that a similar procedure can also be done for spinors in general, where the unitary gauge is the one in which a spinor acquires its polar form, and all non-physical components of the spinor field are transferred into the $P_{\nu}$ and $R_{ij\nu}$ tensors. We have seen also that this procedure can be done on spinor fields that are singular as well, but for this case there are two sub-cases of interest where the $P_{\nu}$ and $R_{ij\nu}$ tensors absorb all components and so that these spinors remain with no true degree of freedom left. Such instances of spinors which are pure Goldstone modes are found to be the well-known Majorana and Weyl spinors.

Weyl and Majorana spinors have peculiar features with respect to their chirality because they are the spinors for which either a single chiral part exists or both chiral parts exist but are complex conjugated of one another. Either way, the full spinor has a single chiral part that is truly independent. This however makes the spinor mathematically equivalent to the doublet of complex scalar fields given by the Higgs field in the standard model, for which transformation properties can always be used to remove all degrees of freedom but one. Nevertheless, for spinors the extra boosts allow the removal of the remaining degree of freedom leaving Majorana and Weyl spinors with no proper intrinsic character. In essence, Weyl and Majorana spinors have only components that can be transferred into frames, and so they are pure Goldstone states.

What lesson we can learn from this? Apart from pure mathematical interest, the physical meaning of the above results is that Weyl and Majorana spinors behave exactly in the same way in which the Higgs would if it were possible to transfer all $4$ components into the gauge fields, that is there would be $4$ massive vector fields but no physical degree of freedom that would remain in the Higgs itself.

This fact seems to point toward the situation for which, like we would consider the Higgs unphysical in the above-mentioned case, we should consider Majorana and Weyl spinors as unphysical too. Perhaps this can indicate that physics should not be constructed on them, or on singular spinor fields, but only on the regular Dirac fields alone.

Yet another interpretation might be that Weyl and Majorana spinors should simply be regarded as topological defects intrinsic to the space-time structure as a whole.

We will leave this question to further investigation.

\ 

Manuscript has associated data in a repository.

\

There is no conflict of interest.

\end{document}